\documentclass[twocolumn,prb,superscriptaddress]{revtex4-1}

\usepackage{hyperref} 
\usepackage{amssymb}
\usepackage{amsmath}
\usepackage{graphicx}

\newcommand{\Div}[1]{\mathbf{\nabla}\cdot #1}
\newcommand{\V}[1]{\mathbf{#1}}
\newcommand{\grad}[1]{\mathbf{\nabla} \left(#1\right)}
\newcommand{\rot}[1]{\mathbf{\nabla}\times #1}

\begin{document}

\title{The impact of nonlocal response on metallo-dielectric multilayers and optical patch antennas}

\author{A. Moreau}
\affiliation{Center for Metamaterials and
  Integrated Plasmonics, Duke University, Durham, North Carolina
  27708, USA} 
\affiliation{Clermont Universit\'e, Universit\'e Blaise
  Pascal, Institut Pascal, BP 10448, F-63000 Clermont-Ferrand, France}
\affiliation{CNRS, UMR 6602, IP, F-63171 Aubi\`ere, France} 
\author{C. Cirac\`i}
\author{D. R. Smith}
\affiliation{Center for Metamaterials and
  Integrated Plasmonics, Duke University, Durham, North Carolina
  27708, USA} 

\begin{abstract}
We analyze the impact of nonlocality on the waveguide modes of
metallo-dielectric multilayers and optical patch antennas, the latter
formed from metal strips closely spaced above a metallic plane. We
model both the nonlocal effects associated with the conduction
electrons of the metal, as well as the previously overlooked response
of bound electrons.  We show that the fundamental mode of a
metal-dielectric-metal waveguide, sometimes called the gap-plasmon, is
very sensitive to nonlocality when the insulating, dielectric layers
are thinner than 5 nm. We suggest that optical patch antennas, which
can easily be fabricated with controlled dielectric spacer layers and
can be interrogated using far-field scattering, can enable the
measurement of nonlocality in metals with good accuracy.
\end{abstract}

\maketitle

\section{Introduction}
With the emergence of new analytical, numerical and nanofabrication
tools, the pursuit of plasmonic systems for a variety of nanophotonic
applications has expanded rapidly in recent
years\cite{maier07,gramotnev2010plasmonics,schuller2010plasmonics}. Plasmonic
media here can be defined as conducting surfaces and nanostructures,
whose optical scattering is largely dominated by the response of the
conduction electrons. Plasmonic behavior is typically associated with
excitation wavelengths at which the inertial inductance of the charge
carriers plays a critical role in the collective
response\cite{zhou05}. In the design of plasmonic media, the dynamics
of the conduction electrons can often be well approximated by assuming
a Drude-like model for the permittivity, which has the frequency
dispersive form
\begin{equation}
\epsilon= 1- \frac{\omega_p^2}{\omega^2+i\Gamma \omega}\label{drude}
\end{equation}
assuming a time dependence of $e^{-i\omega\,t}$. The plasma frequency, $\omega_p$,
proportional to the square root of the carrier density, typically lies
within the ultraviolet portion of the spectrum for many metals. Thus,
for frequencies just below the plasma frequency, the electric
permittivity can be characterized as a lossy dielectric, for which the
real part of the permittivity is moderately negative. At wavelengths
where the real part of the permittivity is negative, surface plasmon
modes can be supported, which are collective oscillations of the
coupled electromagnetic field and conduction electrons. Surface
plasmons can serve to transport energy along metal surfaces in a
manner similar to dielectric waveguides, but are also playing an
increasingly important role in the field of metamaterials, where
metallic nanostructures are often used as elements that provide strong
and customizable scattering. Surface plasmons represent the
underlying mechanism behind perfect lenses\cite{pendry2000negative,fang05}, hyperlenses\cite{jacob2006optical,liu2007far},
spasers\cite{bergman03,zheludev08,oulton09} and many other proposed metamaterial-related devices.

The simplicity of the Drude model of electron response, \eqref{drude}, has
has enabled the rapid modeling of plasmonic and metamaterial structures;
The salient features associated with most plasmonic structures
presented to date can usually be computed with sufficient
accuracy --sometimes even analytically-- assuming the Drude
formula. Particularly when the underlying physics is the main focus
rather than detailed performance characteristics, Eq. \eqref{drude} frequently
provides an adequate description of the plasmonic response. It should
be noted that despite the relatively simple form of Eq. \eqref{drude}, the
numerical simulation of plasmonic systems remains a non-trivial task
because the surface plasmon spatial variation is not limited by the
wavelength of light; rather, the surface plasmon can confine light to
nanometer sized regions, making plasmonic structures an inherently
multiscale modeling problem\cite{kottmann00}. Thus, the frequency dispersion and the negative
permittivity associated with the Drude model contain non-trivial
physics, and have been successfully applied to a wide range of
plasmonic and metamaterial configurations. Naturally, the actual electronic response of a metal or highly doped
semiconductor is much more complicated than that suggested by
Eq. (1). Plasmonic structures are now reliably fabricated at the
nanometer and sub-nanometer scales, where new optical properties arise
that cannot be accounted for solely by the Drude model\cite{scholl12,fernandez12}. Since
these sub-nanometer features are likely to be crucial for optimizing
field localization and enhancement\cite{pendry12,ciraci12,moreau12}, a more
detailed description of the properties of plasmonic devices is
demanded. Effects that would be secondary or of no consequence to the
overall function of prior plasmonic devices, may introduce major
constraints on the detailed performance and ultimate competitiveness
of optimized plasmonic structures with subnanometer features. For
these reasons, it is relevant to consider a more advanced physical
model of the carrier response in conductors.

The Drude model of a conductor assumes only the participation of
conduction electrons (no bound charges), and further assumes a
straightforward force-response relationship between the applied
electric field and responding current density. An intrinsic feature of
this model is that the responding current density at a given point
within the material is proportional to the electric field at that
point; that is, the Drude model assumes locality. Even when the
exact, measured values of the bulk permittivity are used, there is an
implicit assumption of locality since the permittivity is only a
function of frequency rather than of both frequency and wave
vector. To capture the additional physics associated with electronic
response, it is necessary to consider a more detailed model of the
force-response relationship between the field and current density.

More accurate descriptions of the free electron gas have been proposed
in the past, including a description based on a hydrodynamical model
for the conduction
electrons\cite{kliewer68,barrera81,boardman82,gerhardts84,forstmann86,
  fuchs87,fuchs1981dynamical,scalora10}, and a microscopic description
initiated by
Feibelman\cite{feibelman82,liebsch1997electronic,wang11}. The latter
has been improved over the years\cite{liebsch1995influence} and has
been recently used to include the effects of nonlocality on metallic
slabs\cite{ruppin05b} and slot waveguides\cite{wang07}. The
hydrodynamical approach clearly suffers from an uncertainty about
which additional boundary conditions should be used, but allows for
more transparent physical interpretations\cite{fuchs1981dynamical}.
Moreover, the hydrodynamical model can be reasonably implemented in
numerical calculations, and also is useful for finding closed-form,
analytical results. For example, the hydrodynamical model has been
used in conjunction with transformation optics techniques to find
analytical expressions for nanostructures that illustrate the impact
of nonlocal response\cite{pendry12,fernandez12}. Recent experiments
have shown that the hydrodynamical model is able to describe very
accurately the plasmon resonance shift exhibited by spherical
nanoparticles interacting with a metallic film\cite{ciraci12}.
While the hydrodynamical model is clearly not the most sophisticated
approach to describe the free electron gas, it can obviously capture
the physics of nonlocality and it seems it can be made quite accurate
through a correct choice of the free parameters it contains for
situations of interest in plasmonics.

In the present work, we first try to describe the response of the
bound electrons as a polarizable medium, as has been shown to be
accurate for Feibelman's method\cite{liebsch1995influence}, but in the
framework of the hydrodynamical model. We find that this description
greatly simplifies the discussion with respect to the additional
boundary conditions. We explore the consequences of the model on the
reflection of a wave by a metallic surface, on the surface plasmon and
finally on the propagation of a guided wave along a thin metallic
waveguide, as Wang and Kempa have shown that nonlocal effects could be
expected\cite{wang07} for such a structure. Using an analytical
dispersion relation, we show that the nonlocal effects are enhanced in
the slow light regime, when the waveguide is a few nanometers
thick. Finally we study the large impact of nonlocality on optical
patch nanoantennas\cite{miyazaki2006controlled,bozhevolnyi07,
  sondergaard08,jung09,yang12} where the gap beneath the patch behaves as a
cavity, making these structures extremely
sensitive\cite{moreau12}. The optical patch geometry paves the way for
future experiments in which the effects associated with nonlocality
will have easily measurable effects at wavelengths in the visible.

\section{Nonlocal response of metals\label{nonlocal}}

\begin{figure}[h]
\begin{center}
\includegraphics[width=8cm]{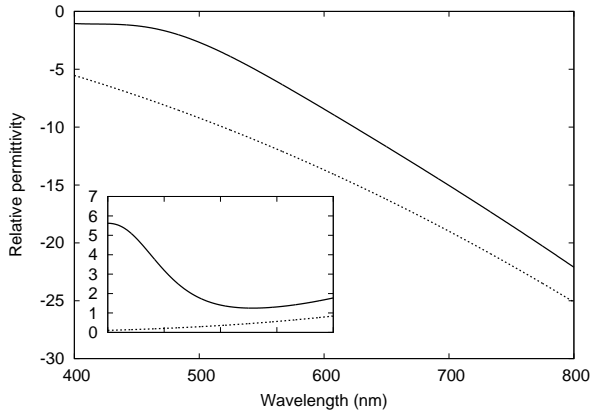}
\end{center}
\caption{Real part of the relative permittivity of gold in the visible
  (solid line) and Drude permittivity according to
  \cite{rakic98} (dotted line). The difference is the contribution of bound electrons. 
  Inset : imaginary part of the relative
  permittivity, same $x$ scale, same lines.\label{fig:eps}}
\end{figure}

While our analysis is not specific to metals, we use the term metal
throughout while keeping in mind the analysis can be applied to highly
doped semiconductors\cite{hoffman07} and potentially other conducting
systems\cite{tassin12}. The polarization of a metal, and hence its
dielectric function, generally contains contributions from both bound
and free conduction electrons. Because we need to apply different
physical response models to the free and bound electrons, it is
essential to first distinguish their relative contributions. The
experimental permittivity curves can be fit\cite{rakic98} with a Drude
term \eqref{drude} that models the free electron contribution, to
which is added a sum over the
Brendel-Bormann\cite{brendel1992infrared} oscillator terms that models
the susceptibility arising from the bound electron
contributions. Figure \ref{fig:eps} shows the permittivity of gold
obtained through the model as well as the fitted Drude permittivity,
corresponding to $1+\chi_f$, where $\chi_f$ is the the susceptibility
of the free electrons. The difference (not shown) between the modeled
permittivity and the fitted Drude term corresponds to the contribution
of the bound electrons, $\chi_b$.

We assume here that the nonlocal response of the metal is largely
dominated by the nonlocality induced by free electrons, so that 
we can treat the bound electron contribution as purely local,
as some authors do\cite{mcmahon09,mcmahon10,fernandez12}.
Bound electrons too can be expected to present a nonlocal response,
similar to what occurs in dielectrics\cite{maradudin73,agranovich}.
Our assumption is equivalent to assuming that the interactions between
electrons in a free electron gas (through a quantum pressure and
Coulomb repulsion) are more intense than essentially dipole-dipole
interaction between bound electrons.

Under an applied electric field, the medium will undergo a
polarization with contributions from both bound and free
electrons. The total polarization vector can thus be written
\begin{equation}
\V{P}= \V{P}_b + \V{P}_f
\end{equation}
where $\V{P}_b =\epsilon_0 \chi_b \V{E}$, $\chi_b$ being the
susceptibility of the bound electrons, with the currents in the free
electron gas related to the polarization in the usual manner:
\begin{equation}
\dot{\V{P}}_f=\V{J}.
\end{equation}

By incorporating all responding currents and charges into the
polarization, we can treat the metal as a dielectric, such that the
electric flux density, $\V{D} = \epsilon_0 \V{E} + \V{P}_b + \V{P}_f$,
satisfies $\Div{\V{D}}=0$. Taking the divergence of $\V{D}$ and
writing $\V{P}_b$ in terms of the electric field, we obtain
\begin{equation}
\Div{\V{P}_f}=-\epsilon_0 (1+\chi_b)\,\Div{\V{E}}.\label{eq:div}
\end{equation}
where we have explicitly assumed that the bound electron
susceptibility is local and can be taken outside the divergence
operator.

The free electron current density can be related to the applied
electric field using the hydrodynamical model. Using Eq. 2, a
linearized equation relating $\V{P}_f$ to the electric field is given
by\cite{scalora10}
\begin{equation}
-\beta^2\,\grad{\Div \V{P}_f}+\ddot{\V{P}}_f+\gamma \dot{\V{P}}_f =
\epsilon_0 \omega_p^2 \V{E} \label{p2}
\end{equation}
where $\gamma$ is the damping factor, due to collisions of the
electron gas with the ion grid, $\omega_p$ is the plasma frequency of
the metal, and $\beta$ is the phenomenological nonlocal parameter,
proportional to the Fermi velocity $v_F$. Usually the value of
$\beta=\sqrt{\frac{2}{3}}v_F$ has been considered in the
literature. However, a slightly more realistic hydrodynamic model
should take into account other sources of nonlocality, such as the
Bohm potential, which can be shown to be of the same order of the
Fermi pressure\cite{crouseilles08}. Though it is beyond the scope of this
paper to introduce a more sophisticated model, it makes sense from a
phenomenological approach to consider a more empirical value for the
parameter $\beta$. Recently it has been shown for plasmonic systems of
film-coupled gold nanoparticles that the value $\beta=
\sqrt{\frac{5}{3}\frac{E_F}{m}}\simeq 1.27 \times 10^6~\rm{m/s}$ gives
a very good agreement with experimental data\cite{ciraci12}. In this
work we will then assume this former value for both gold and silver.

Assuming a harmonic solution of the form $e^{-i\omega t}$, and using equation \ref{eq:div}, the polarization $\V{P}_f$ can finally be
written
\begin{equation}
\V{P}_f = - \epsilon_0 \frac{\omega_p^2}{\omega^2+i\gamma\omega} \left(\V{E} - (1+\chi_b)\,\frac{\beta^2}{\omega_p^2} \,\grad{\Div \V{E}} \right)\label{eq:p2},
\end{equation}
where the term
\begin{equation}
\chi_f = -\frac{\omega_p^2}{\omega^2+i\gamma\omega}
\end{equation} 
can be identified as the local susceptibility associated with free electrons, corresponding to the Drude model.

We have written the polarization terms in such a manner that the free
and bound electron contributions can be distinguished. In determining
the various parameters in these equations for the calculations that
follow, we use the model provided by\cite{rakic98} and shown Fig. 1.

\section{Transverse and longitudinal modes in metals\label{derivation}}

In a metal, taking the above description of nonlocality into account, Maxwell's equations can now be written
\begin{align}
\rot{\V{E}} &=i\omega \mu_0 \V{H}\label{eq:M1}\\
\rot{\V{H}}&=-i\omega\left(\epsilon_0 (1+\chi_b)\V{E} + \V{P}_f\right)\\
&=-i\omega\epsilon_0\epsilon \left(\V{E}-\alpha \grad{\Div{\V{E}}}\right)\label{eq:M2}
\end{align}
where $\epsilon$ is the local relative permittivity of the metal
\begin{equation}
\epsilon=1+\chi_b +\chi_f
\end{equation}
and
\begin{align}
\alpha &=\frac{\chi_f\,(1+\chi_b)}{\epsilon}\,\frac{\beta^2}{\omega_p^2}\\
&= \frac{\beta^2}{\frac{\omega_p^2}{1+\chi_b} - \omega^2 - i\gamma\omega}.
\end{align}

As shown rigorously in the appendix, there are two different solutions
to these equations corresponding to two different kinds of waves. The
first solution satisfies $\Div{\V{E}}=0$, so that it corresponds to
the standard solution to Maxwell's equations when the nonlocality is
overlooked. Equations \eqref{eq:M1} and \eqref{eq:M2} become
\begin{align}
\rot{\V{E}} &= i\omega \mu_0 \V{H}\\
\rot{\V{H}} &= -i\omega \epsilon_0 \epsilon \V{E},
\end{align}
Finally all the fields satisfy Helmholtz's equation
\begin{equation}
\nabla^2 \V{H} +\epsilon\,k_0^2 \V{H}=\V{0}\label{eq:helmholtz},
\end{equation}
where $k_0=\frac{\omega^2}{c^2}$. Since the divergence of the electric
field is zero, the electric field is orthogonal to the wavevector when the
wave is propagative, which means it is transverse. The dispersion
relation for these transverse waves is thus
\begin{equation}
\V{k}^2=\epsilon\,k_0^2 = \epsilon\,\frac{\omega^2}{c^2}\label{eq:transverse}.
\end{equation}

The second kind of solution is curl free, which means it satisfies
$\rot{\V{E}}=\V{0}$ and there is no accompanying magnetic field. These
waves are called longitudinal because when they are propagative, the electric field is parallel to the
wavevector. They correspond to bulk plasmons: oscillations of the
free electron gas due to the pressure term. Since the divergence of
the electric field is not identically zero, there exists a charge density inside the metal given by
\begin{equation}
\rho = \epsilon_0 \Div{\V{E}}.
\end{equation}
Equation \eqref{eq:M2}, then yields the wave equation for the bulk plasmons
\begin{equation}
\grad{\Div{\V{E}}}-\frac{1}{\alpha}\V{E} = \nabla^2 \V{E}-\frac{1}{\alpha}\V{E} = \V{0}.\label{eq:bulk}
\end{equation}
and the corresponding dispersion relation is 
\begin{equation}
\V{k}^2 = -\frac{1}{\alpha} =\frac{1}{\beta^2}\left(\omega^2 -\frac{\omega_p^2}{1+\chi_b}+ i\gamma\omega\right)\label{eq:LW}.
\end{equation}

An alternative way to write this dispersion relation is 
\begin{equation}
\epsilon_{\parallel}\equiv1+\chi_b - \frac{\omega_p^2}{\omega^2+i\gamma\omega - \beta^2 \V{k}^2} = 0\label{eq:e//},
\end{equation}
which is the way previous works have taken $\chi_b$ into
account\cite{fernandez12} through a so-called 
longitudinal permittivity. But the
equation governing the polarization $\V{P}_f$ (equation \eqref{p2})
cannot be deduced from the longitudinal permittivity
using a simple Fourier transform\cite{mcmahon09,mcmahon10}, as has been previously pointed out\cite{raza11}.

\begin{figure}[h]
\begin{center}
\includegraphics[width=8cm]{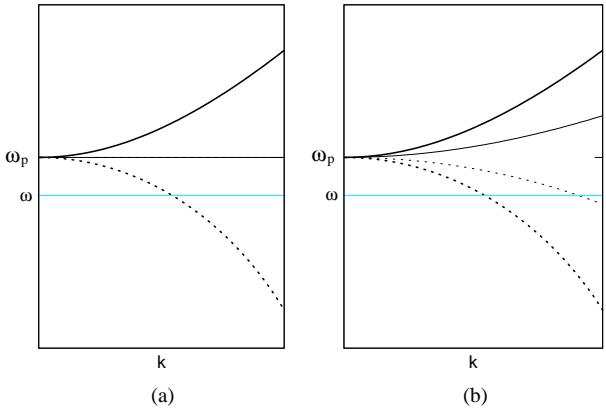}
\end{center}
\caption{Dispersion relation for transverse waves (thick lines) and bulk plasmons (thin lines)
when (a) the nonlocality is absent ($\beta=0$) and (b) the nonlocality is important. The dashed
lines show the imaginary parts of the propagation constants for the transverse and longitudinal
waves, below the plasma frequency $\omega_p$. When nonlocality is present, for a given $\omega$
two waves must be taken into account.(color online)\label{fig:disp}}
\end{figure}

The dispersion relations, Eqs. \eqref{eq:transverse} and \eqref{eq:LW}, are plotted in Fig. \ref{fig:disp}
for two cases of the nonlocal parameter $\beta$, for the simplified case where $\epsilon = 1-\frac{\omega_p^2}{\omega^2}$. 
For small $\beta$, the longitudinal mode disperses very little, and can be generally ignored
in wave propagation problems. When $\beta$ is nonzero, however, the
longitudinal mode acquires dispersion, and is generally present at a
given frequency of excitation. Above the plasma frequency, both the
transverse and longitudinal modes are propagating, while below the
plasma frequency both modes decay exponentially. In considering
boundary value problems, it is clear that a wave incident on a half
space filled with a nonlocal, plasmonic medium will generally couple
to both types of waves. To avoid the system being underdetermined, an
additional boundary condition must be used as will be discussed
in the subsequent section.

\begin{figure}[h]
\begin{center}
\includegraphics[width=8cm]{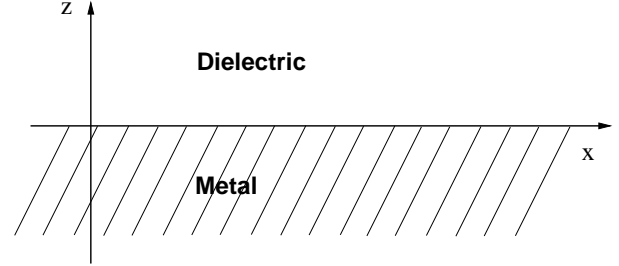}
\end{center}
\caption{A simple interface between a dielectric and a metal.\label{fig:schema0}}
\end{figure}

Let us now consider a multilayered structure, that could be as simple
as the single interface shown in figure \ref{fig:schema0}, invariant in
two directions, here taken as $x$ and $y$. The $z$ axis is thus
perpendicular to any interface considered, as shown in Fig.
\ref{fig:schema0}. Without any loss of generality, it is possible to
assume solutions that are translationally invariant along the $y$
(out-of-plane) direction. As shown in the appendix, the system of
equations \eqref{eq:M1} and \eqref{eq:M2} can be split into two
subsystems corresponding to $s$ (electric field polarized
perpendicular to the plane of incidence) and $p$ (magnetic field
polarized perpendicular to the plane of incidence) polarizations.
Moreover, we will assume from now on that all the fields present an
$x$ dependence that varies as $e^{ik_x\,x}$ (or, equivalently, we take
the Fourier transform along the $x$ axis).

For the $s$ polarization, $\rot{\V{E}}=\V{0}$ yields $E_y=0$, so that
no bulk plasmon can be excited. Nonlocality has then no impact on this
polarization, so that we will deal in the following with $p$
polarization only.

Equation \eqref{eq:helmholtz} then yields
\begin{equation}
\partial_z^2 H_y = -(\epsilon\,k_0^2-k_x^2)\,H_y,
\end{equation}
so that the magnetic field can be written
\begin{equation}
H_y = (A\,e^{i{k_z}_t\,z}+B\,e^{-i{k_z}_t\,z})\,e^{ik_x\,x-i\omega\,t}
\end{equation}
with ${k_z}_t =\sqrt{\epsilon\, k_0^2 -k_x^2}$ where $k_0 = \frac{\omega}{c}$.
The $E_x$ and $E_z$ accompanying fields can be found using equations 
\begin{align}
E_x &= \frac{1}{i\omega \epsilon_0\,\epsilon}\partial_z H_y\label{eq:Ex}\\
E_z &=-\frac{1}{i\omega \epsilon_0\,\epsilon}\partial_x H_y\label{eq:Ez}.
\end{align}

For longitudinal waves, the wave equation \eqref{eq:bulk} becomes
\begin{equation}
\partial_z^2 E_x = \left(k_x^2+\frac{1}{\alpha}\right)\,E_x \,.
\end{equation}
For normal incidence (for $k_x = 0$), depending whether $\omega$ is smaller or larger than $\frac{\omega_p}{\sqrt{1+\chi_b}}$, the bulk plasmon will be
respectively evanescent ($\Re (\alpha) >0$) or propagative ($\Re (\alpha)
< 0$) . In the visible range, we usually
have $\omega<\frac{\omega_p}{\sqrt{1+\chi_b}}$ so that the bulk plasmon is evanescent and the above equation can be solved to yield
\begin{equation}
E_x = (C\,e^{\kappa_l\,z}+D\,e^{-\kappa_l\,z})\,e^{ik_x\,x-i\omega\,t}
\end{equation}
with 
\begin{align}
\kappa_l^2 &= \frac{1}{\beta^2}\left(\frac{\omega_p^2}{1+\chi_b}-\omega^2 -i\gamma\omega\right)+k_x^2\\
&= k_x^2 + \frac{\omega_p^2}{\beta^2}\left(\frac{1}{\chi_f}+\frac{1}{1+\chi_b}\right).
\end{align}

The fact that longitudinal waves are curl-free yields 
\begin{equation}
E_z = \frac{1}{ik_x}\partial_z E_x,
\end{equation}
which allows determination of the contribution of the bulk plasmon to $E_z$ if needed.

\section{Additional boundary conditions\label{abc}}

The nonlocal nature of the metal results in the appearance of a
longitudinal bulk plasmon mode that can be excited from the metal
interface, in addition to the surface-localized plasmon polariton.
The well-known Maxwell's boundary conditions are not sufficient to
uniquely define the amplitudes of these independent waves.  More
specifically, for each metallic layer, two new unknowns are introduced
and must be resolved in the solution of the electromagnetic boundary
value problem.  To avoid dealing with an underdetermined problem then,
\textit{additional boundary conditions} must be imposed at the metal
interface.
 
The issue of boundary conditions has been abundantly discussed in the
context of spatially dispersive crystals, and a variety of different
boundary conditions has been proposed\cite{agranovich,halevi84}.  In
the context of the hydrodynamic model, the choice of boundary
conditions is much simpler\cite{boardman82,forstmann86}, essentially
because fewer types of waves are involved. Two additional boundary
conditions are typically considered in the case of an interface
between a metal and a dielectric, when the contribution of the bound
electrons is overlooked: either (i)
$P_{z}=0$\cite{boardman82,forstmann86} or (ii) the continuity of
$E_z$\cite{ruppin05,fernandez12}. If the considered dielectric is
vacuum, then these two conditions are equivalent. Condition (i) can be
justified because the polarization in the metal is due to actual
currents; since the electrons are not allowed to leave the metal, then
the normal current must vanish at the interface and also the
polarization. Condition (ii) can be justified by treating the
interface as smooth for all fields, including the normal component of
the electric field.

In our description of the response of metals, the susceptibility
attributed to bound electrons, $\chi_b$ is considered purely local.
One might expect that the equation $\Div \V{D} = 0$ would impose a
supplementary condition (namely the continuity of $D_z$), leaving no
freedom in the choice of the boundary condition. This is however not
the case: in multilayered systems, the continuity of $H_y$ through an
interface implies the continuity of $D_z$, so that an additional
boundary condition is still required.

The response of the metal in our description is partly the response of
a standard dielectric medium, so that there is no reason to assume the
continuity of $E_z$ at the surface of the metal. Condition (ii) thus
appears very difficult to support when the contribution of bound
electrons is taken into account as a local, polarizable medium.

The underlying physics\cite{boardman82,forstmann86} behind condition
(i), that free electrons cannot escape the metal, does however not
lead here to $P_z=0$ at the edge of the metal because not all the
polarization comes from actual currents in the free electron gas. It
is thus not reasonable to use boundary condition (i) for the case when
bound electrons contribute to the polarization response.

It would be physically reasonable to consider that only the
polarization linked to actual current leaving the metal should be zero
at the interface between a metal and a dielectric. For multilayered
structures, this condition can be written
\begin{equation}
{P_{f}}_z = 0.
\end{equation}
at the interface as an additional boundary condition.
We underscore that this boundary condition is {\em not} equivalent to
conditions (i) and (ii) when the outside medium is vacuum.

In the case of an interface between two metals, again, condition (i)
is hard to justify, but the interface obviously should not be
considered as impervious to free electrons.  Instead, it would sound to
consider that the currents, and thus the polarization $\V{P}_f$,
should be continuous. This would actually provide the two additional
boundary conditions that are required for an interface between two
metals. Although we will not consider here structures involving such
an interface, we emphasize that taking into account the contribution
of bound electrons to the response of metals seem to lead to unambiguous
boundary conditions based on physical reasoning.

\section{Reflection from a metallic surface\label{reflection}}

Let us now consider an incident plane wave coming from above ($z>0$) and
propagating in a dielectric medium with a permittivity $\epsilon_d$,
reflected by a metallic interface located at $z=0$, as shown
in Fig. \ref{fig:schema0} - the metal being characterized by a
permittivity $\epsilon$. 

For $p$ polarization, the magnetic field in the dielectric region
can be then written
\begin{equation}
H_y = \left(
e^{-ik_z\,z} + r\, e^{ik_z\,z}
\right) \,e^{ik_x\,x-i\omega\,t}
\end{equation}
where $k_z = \sqrt{\epsilon_d\,k_0^2 - k_x^2}$ and $k_0 = \frac{\omega}{c}$, 
while the electric field along the $x$ direction has the form
\begin{equation}
E_x=\frac{ik_z}{i\omega\epsilon_0\epsilon_d}\,\left(r\, e^{ik_z\,z}-e^{-ik_z\,z}\right)\,e^{ik_x\,x-i\omega\,t}.
\end{equation}
In the metal, the magnetic field can be written
\begin{equation}
H_y = A \,e^{\kappa_t\,z}\,e^{ik_x\,x-i\omega\,t}
\end{equation}
where $\kappa_t = \sqrt{k_x^2-\epsilon\,k_0^2}$, and the electric field
\begin{align}
E_x &=\left(\frac{\kappa_t}{i\omega\epsilon_0\epsilon} A \,e^{\kappa_t\,z}
+B\,e^{\kappa_l\,z}
\right)\,e^{ik_x\,x-i\omega\,t}\\
E_z &=\left(-\frac{ik_x}{i\omega\epsilon_0\epsilon} A \,e^{\kappa_t\,z}
+\frac{\kappa_l}{ik_x}\,B\,e^{\kappa_l\,z}
\right)\,e^{ik_x\,x-i\omega\,t}.
\end{align}

The magnetic field $H_y$ and the $x$ component of the electric field $E_x$ are continuous at $z=0$ so that
\begin{align}
1+r &= A\\
(r-1) \frac{ik_z}{\epsilon_d} &= \frac{\kappa_t}{\epsilon}\,A + i\omega\epsilon_0 \,B.
\end{align}
Since ${P_{f}}_z = -\frac{1}{i\omega} \,\partial_x H_y -\epsilon_0 (1+\chi_b)\,E_z$, the condition ${P_{f}}_z=0$
in the metal at the interface, can be written
\begin{equation}
ik_x\,A\,\left(
\frac{1}{\epsilon} - \frac{1}{1+\chi_b}
\right)
=\frac{\kappa_l}{ik_x}\,i\omega\epsilon_0 \,B
\end{equation}

Finally $A$ and $B$ can be eliminated to yield
\begin{equation}
r = \frac{
\frac{ik_z}{\epsilon_d} + \frac{\kappa_t}{\epsilon} -  \Omega
}{
\frac{ik_z}{\epsilon_d} - \frac{\kappa_t}{\epsilon} + \Omega
}\label{eq:r}
\end{equation}
where 
\begin{equation}
\Omega = \frac{k_x^2}{\kappa_l}
\left(
\frac{1}{\epsilon}
-
\frac{1}{1+\chi_b}
\right)\label{eq:Omega}
\end{equation}

The reflection coefficient indicates that the bulk plasmon is not
excited at normal incidence for $k_x=0$ because in that case only one
component of the electric field is present in the incident and
reflected fields. When the angle of incidence increases the excitation
of the bulk plasmon is more and more important because of the
increasing $E_z$ component. Of course $\kappa_l$ is increasing too,
which means that the bulk plasmon penetration is more shallow, but
only slightly - so that the $\Omega$ increases essentially as $k_x^2$.

\section{Surface plasmon\label{sp}}

If the field is not propagative in the dielectric region, but has the form
\begin{equation}
H_y = \left(
C\,e^{\kappa_z\,z} + D\,e^{-\kappa_z\,z}
\right) \,e^{ik_x\,x-i\omega\,t}
\end{equation}
with $\kappa_z = \sqrt{k_x^2 - \epsilon_d \,k_0^2} = -ik_z$, then it is meaningless to define a 
reflection coefficient \eqref{eq:r}, but still we can write that
\begin{equation}
\frac{D}{C} = 
\frac{
\frac{\kappa_z}{\epsilon_d} - \frac{\kappa_t}{\epsilon} +  \Omega
}{
\frac{\kappa_z}{\epsilon_d} + \frac{\kappa_t}{\epsilon} - \Omega
}\label{eq:r2}
\end{equation}

The surface plasmon is a solution for which $D\neq 0$ and $C=0$, thus corresponding to
a pole of the left hand side of equation \eqref{eq:r2}, and a zero of its denominator, so
that the dispersion relation can be written
\begin{equation}
\frac{\kappa_z}{\epsilon_d} + \frac{\kappa_t}{\epsilon} = \Omega
\end{equation}

\begin{figure}[h]
\begin{center}
\includegraphics[width=8cm]{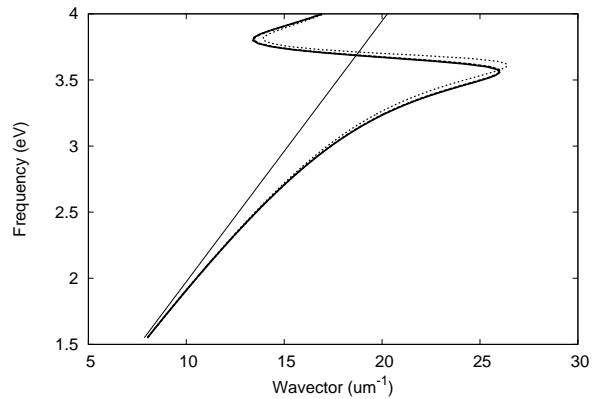}
\end{center}
\caption{Dispersion relation ($\omega$ as a function of $k_x$) for a surface plasmon at the interface between silver and air.
The local description (thick solid curve) can almost not be distinguished from the nonlocal description (thick dashed curve, almost identical with the thick solid curve). In order to illustrate the effect of nonlocality, we show here (dotted line) the impact of an exaggerated nonlocality ($\beta$ multiplied by ten). The thin solid curve is $\omega=k_x\,c$.\label{fig:sp}}
\end{figure}

The larger the propagation constant $k_x$, the larger $\Omega$ and
thus the larger the impact of nonlocality. However, for surface
plasmons, very large values of $k_x$ are difficult to reach
(typically, the maximum effective index is around 1.4 for a silver-air
interface) as shown in figure \ref{fig:sp}. The impact of nonlocality
on bare surface plasmons is thus very small. In the following, we will
see that for a metal-dielectric-metal waveguide with a very thin
dielectric layer, the impact of nonlocality on the guided mode is much
more important because very large $k_x$ values can be reached whatever
the wavelength.

\section{The impact of boundary conditions\label{others}}

The form of the reflection coefficient and to the dispersion relation
above clearly show that $\Omega$ is the parameter controlling the
influence of the nonlocality on propagation phenomena. Moreover, it
makes manifest the consequences of a change in the boundary
conditions.

In the literature, the entire metal response is often attributed to
the free electrons, while the response of bound electrons is
neglected\cite{boardman82,forstmann86,ruppin05}. When the bound
electron response is neglected, the dispersion relation of the bulk
plasmon yields
\begin{equation}
\kappa_l= \sqrt{k_x^2 + \frac{\omega_p^2}{\beta^2}\left(\frac{1}{\chi_f}+1\right)}
\end{equation}
instead of 
\begin{equation}
\kappa_l= \sqrt{k_x^2 + \frac{\omega_p^2}{\beta^2}\left(\frac{1}{\chi_f}+\frac{1}{1+\chi_b}\right)}
\end{equation}
where a local contribution from bound electrons is assumed\cite{fernandez12}.

If the boundary condition with the dielectric is chosen to be the
continuity of the component of the electric field normal to the
interface, then we have
\begin{equation}
\Omega = \frac{k_x^2}{\kappa_l}
\left(
\frac{1}{\epsilon}
-
\frac{1}{\epsilon_d}
\right),
\end{equation}
where $\kappa_l$ can be calculated using one of the above expressions,
depending on the description of the metal's properties. When the entire polarization $P_z$ is chosen to vanish at the interface, we have instead
\begin{equation}
\Omega = \frac{k_x^2}{\kappa_l}
\left(
\frac{1}{\epsilon}
-
1
\right).
\end{equation}

In the following, we will investigate all the different descriptions
that are presented in table \ref{tab:desc} to show that, even if they
differ regarding the quantitative impact of nonlocality, they all at
least agree qualitatively.

\begin{table}[h]
\begin{center}
\begin{tabular}{cccc}
Descr.& $\kappa_l^2$ & A.B.C. & $\Omega$ \\
\hline
\hline
1 & $k_x^2 + \frac{\omega_p^2}{\beta^2}\left(\frac{1}{\chi_f}+1\right)$ & $P_z(0) = 0$ & $\frac{k_x^2}{\kappa_l}\left(\frac{1}{\epsilon}-1\right)$\\
2 & $k_x^2 + \frac{\omega_p^2}{\beta^2}\left(\frac{1}{\chi_f}+1\right)$ & $E_z$ continuous & $\frac{k_x^2}{\kappa_l}\left(\frac{1}{\epsilon}-\frac{1}{\epsilon_d}\right)$\\
3 & $k_x^2 + \frac{\omega_p^2}{\beta^2}\left(\frac{1}{\chi_f}+\frac{1}{1+\chi_b}\right)$ & $P_z = 0$ & $\frac{k_x^2}{\kappa_l}\left(\frac{1}{\epsilon}-1\right)$ \\
4 & $k_x^2 + \frac{\omega_p^2}{\beta^2}\left(\frac{1}{\chi_f}+\frac{1}{1+\chi_b}\right)$ & $E_z$ continuous & $\frac{k_x^2}{\kappa_l}
\left(\frac{1}{\epsilon}-\frac{1}{\epsilon_d}\right)$\\
5 & $k_x^2 + \frac{\omega_p^2}{\beta^2}\left(\frac{1}{\chi_f}+\frac{1}{1+\chi_b}\right)$ & ${P_f}_z = 0$ & $\frac{k_x^2}{\kappa_l}\left(\frac{1}{\epsilon}-\frac{1}{1+\chi_b}
\right)$\\
\hline
\end{tabular}
\caption{\label{tab:desc}Summary of the different descriptions of
  nonlocality. The first two do not consider separately the
  contribution from the bound electrons, the last three do. The last
  one is the one that is preferred in this work.}
\end{center}
\end{table}

\section{Metal-dielectric-metal waveguide}

While the impact of nonlocality can be considered minor for the single
interface problem above, nonlocal effects can be far more evident in
multilayer systems. In metallo-dielectric layers, it is possible to
reduce the thickness of layers to the nanometer or even sub-nanometer
scale; modes that propagate in such layers can be significantly
confined, to the point where local models are forced to break
down. For this reason, multilayer systems and structures based on
multilayers can be useful as an experimental tool to investigate and
measure nonlocal effects.

\begin{figure}[h]
\begin{center}
\includegraphics[width=8cm]{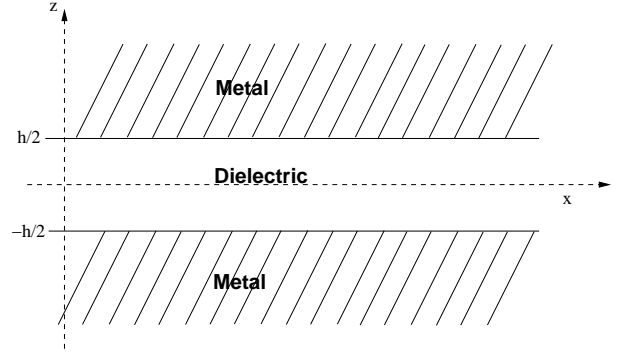}
\end{center}
\caption{Metallic waveguide of width $h$.\label{schema}}
\end{figure}

In this section, we consider the case of a dieletric with a
permittivity $\epsilon_d$ sandwiched between two metallic surfaces (as
shown in figure \ref{schema}) and study more thoroughly the influence
of the nonlocality of the metal on the first even guided mode (the
fundamental mode).

\subsection{Dispersion relation}

We consider here a symmetric waveguide, the metal being the same on
both sides of the dielectric layer. The magnetic field in the
dielectric can be written as
\begin{equation}
H_y = \left( C\,e^{\kappa_z\,z} + D\,e^{-\kappa_z\,z} \right) \,e^{ik_x\,x-i\omega\,t}\,.
\end{equation}

As we have seen in the previous section, at $z=-\frac{h}{2}$, we have
\begin{equation}
\frac{D\,e^{\kappa_z\,\frac{h}{2}}}{C\,e^{-\kappa_z\,\frac{h}{2}}} = 
\frac{
\frac{\kappa_z}{\epsilon_d} - \frac{\kappa_t}{\epsilon} +  \Omega
}{
\frac{\kappa_z}{\epsilon_d} + \frac{\kappa_t}{\epsilon} - \Omega
}
\end{equation}
while for $z=+\frac{h}{2}$ (the $z$ axis has to be reversed, which means $C$ and
$D$ should be exchanged)
\begin{equation}
\frac{C\,e^{\kappa_z\,\frac{h}{2}}}{D\,e^{-\kappa_z\,\frac{h}{2}}} =\frac{
\frac{\kappa_z}{\epsilon_d} - \frac{\kappa_t}{\epsilon} +  \Omega
}{\frac{\kappa_z}{\epsilon_d} + \frac{\kappa_t}{\epsilon} - \Omega
}.
\end{equation}

Combining these two equations, we get 
\begin{equation}
e^{2\kappa_z\,h} = \left(\frac{\frac{\kappa_z}{\epsilon_d} - \frac{\kappa_t}{\epsilon} +  \Omega
}{\frac{\kappa_z}{\epsilon_d} + \frac{\kappa_t}{\epsilon} - \Omega}\right)^2 = r^2
\end{equation}
and finally either the mode is symetrical ($C=D$) and we have $r=e^{\kappa_z\,h}$ which can be written
\begin{equation}
\frac{\kappa_z}{\epsilon_d}\tanh \frac{\kappa_z\,h}{2} + \frac{\kappa_t}{\epsilon} = \Omega\label{eq:sym}
\end{equation}
or the mode is antisymetrical ($C=-D$), which means $r=-e^{\kappa_z\,h}$ and finally 
\begin{equation}
\frac{\kappa_z}{\epsilon_d}\coth \frac{\kappa_z\,h}{2} + \frac{\kappa_t}{\epsilon} = \Omega.
\end{equation}

\subsection{Nature of the guided modes\label{mdm}}

We first discuss the nature of the guided modes in a thin metallic
waveguide.  There are two situations that are clear and for which the
guided modes of the structure have well-posed
definitions\cite{maier07} :
\begin{itemize}
\item The perfect metallic waveguide, which supports a {\em
  fundamental mode} that is flat and that has no cut-off (it is
  supported whatever the thickness of the metallic waveguide). In
  addition, we have analytical expressions for the propagation
  constant and field profile of all the modes. For the fundamental
  mode, we have
\begin{equation}
k_x^2 = \epsilon_d \,k_0^2
\end{equation}
\item The plasmonic ({\em i.e.} wide) metallic waveguide, which supports {\em
  coupled surface plasmons}\cite{maier07}. At a given frequency and
  for a wide enough guide, the even and the odd surface plasmon modes
  present propagation constants that can be arbitrarily close to the
  propagation constant of the surface plasmon
\begin{equation}
k_x = k_0 \,\sqrt{\frac{\epsilon_d\,\epsilon(\omega)}{\epsilon_d+\epsilon(\omega)}},
\end{equation}
even for complex values of $\epsilon(\omega)$.
\end{itemize}

For the case of a thin (a few nanometers) waveguide we seek the best
description to retain for the only guided mode found.

\begin{figure}[h]
\begin{center}
\includegraphics[width=8cm]{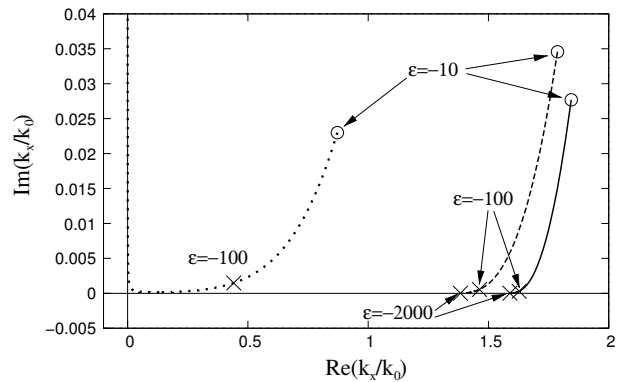}
\end{center}
\caption{Trajectory in the complex plane of the quantity
  $\frac{k_x}{k_0}$ for three different waveguided modes (solid line: first even mode; dashed line: first odd mode ; dotted line: second even mode) when the
  permittivity of the metal goes from $\epsilon=-10+i$ (circles) to
  $\epsilon=-2000+i$, for a thickness of the waveguide of 500 nm and a
  dielectric with an 1.58 optical index. The intermediate value of $-100+i$ 
  is indicated on the curves.\label{fig:e_au}}
\end{figure}

Consider the case of coupled surface plasmons first. We can approach
the condition of a perfect metallic waveguide by making the
permittivity of the metal change such that its real part tends towards
infinity. As can be seen in figure\ref{fig:e_au} the odd mode tends
towards the fundamental mode but the field inside the dielectric
(index of 1.58) always stays evanescent. The even mode tends towards
the first even mode of the perfect metallic waveguide and the field
becomes propagative at some point (where the real part of the
propagation constant becomes smaller than the optical index of the
dielectric). The point at which the field of the even mode becomes
propagative could even be defined as a limit between the ``coupled
surface plasmon'' and the ``perfect metallic waveguide'' pictures.

\begin{figure}[h]
\begin{center}
\includegraphics[width=8cm]{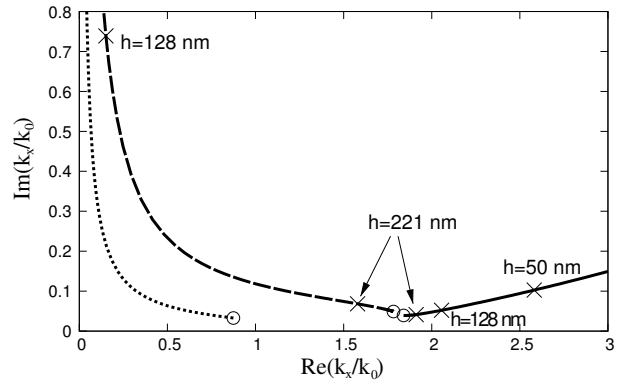}
\end{center}
\caption{Trajectory in the complex plane of the quantity
  $\frac{k_x}{k_0}$ for three different waveguided modes (solide line: first even mode;
  dashed line: first odd mode ; dotted line: second even mode) when the
  width of the metal goes from 380 nm (circles) to 10 nm. The permittivity of
  the metal is taken equal to the permittivity of gold at 608 nm,
  $\epsilon\simeq -10.01+1.44i$. Intermediate thickness of $221$ nm
  (when the real part of the index of the odd mode becomes smaller
  than 1.58, so that the coupled plasmon picture becomes less
  relevant), $128$ nm (when the odd mode can be considered
  non-propagative, and the coupled plasmon picture is not relevant any
  more) and $50$ nm are indicated on the curves.
\label{fig:h}.}
\end{figure}

Now consider starting with a large waveguide (500 nm) and decreasing
its width down to a few nanometers. As can be seen in figure
\ref{fig:h}, the even mode presents an increasing propagation
constant. The odd mode, by contrast, presents a decreasing propagation
constant - the field in the dielectric even becomes propagative as in
the previous case. For thin layers (smaller than 128 nm here) this
mode presents a very large imaginary part and a very small real part:
it can be considered as evanescent in the $x$ direction even if the
cut-off cannot be defined precisely.

For a small dielectric thickness, the waveguide thus behaves much more
like a perfect metallic waveguide (a fundamental mode with no cut-off,
no propagative even mode) and except for the fact that the field of
the first even mode is evanescent in the $z$ direction, has not much
to do with the coupled surface plasmons situation. This is why we
refer to this mode as the {\em fundamental mode} of the waveguide.

This mode is however sometimes called {\em gap-plasmon} in the
literature\cite{jung09}, a term that underscores the differences
between the actual mode and the fundamental more of a {\em perfect}
metallic waveguide.

\subsection{Nonlocal effects\label{thin}}

\begin{figure}[h]
\begin{center}
\includegraphics[width=8cm]{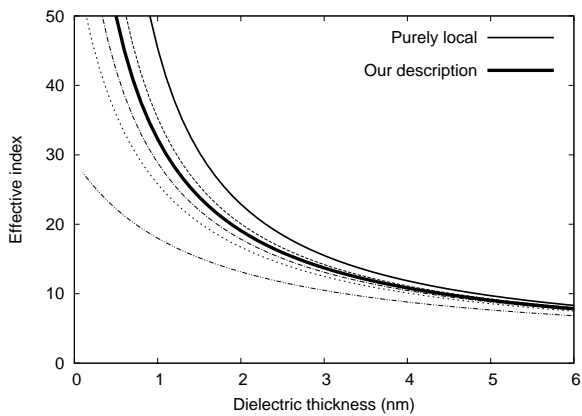}
\end{center}
\caption{Effective index of the guided mode at 600 nm, as a function
  of the dielectric width, $h$. The dispersion relations are shown for
  different descriptions (see table \ref{tab:desc}): the completely
  local case (thin black curve, top), $P_z(0)=0$ (description 1,
  dash-dotted line) and $E_z$ continuous (description 2, dashed line)
  with no identified contribution of the bound electrons, and
  descriptions separating the contributions of bound and free
  electrons, with $P_z=0$ (description 3, dash-double dotted line), a
  continuous $E_z$ (description 4, thin dash line) and finally our
  description (description 5, thick solid curve), that is preferred in
  this work.\label{fig:mode}}
\end{figure}

\begin{figure}[h]
\begin{center}
\includegraphics[width=8cm]{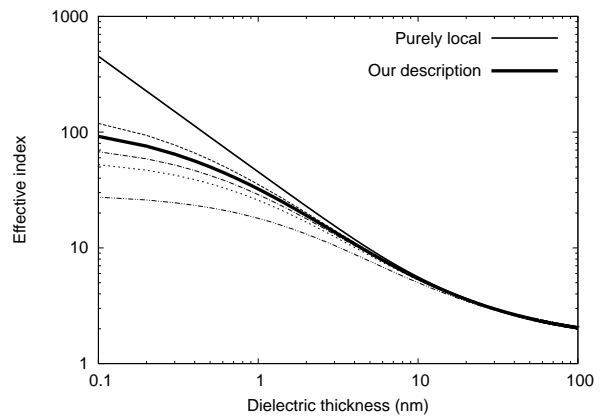}
\end{center}
\caption{Same as figure \ref{fig:mode} except for the log scale.\label{fig:log}}
\end{figure}

When the waveguide becomes extremely thin, as can be seen in figure
\ref{fig:h}, the effective index (and thus $k_x$) of the fundamental
mode (with a dispersion relation given by \eqref{eq:sym}) can become
arbitrary large.  When $k_x$ is larger, $\Omega$ is larger, which
means that the non-locality has a much larger impact on the mode's
propagation constant. It is possible to compare (see figures
\ref{fig:mode}, \ref{fig:log} for a waveguide filled with a dielectric
with a 1.58 optical index) the local effective index as a function of
the waveguide's width with a local and with a nonlocal
theory. Obviously the impact of nonlocality is limited for $h>5$ nm
but it can become very important under that threshold. The parameters
we have considered for gold are given in \cite{rakic98} and
$\beta = 1.27.10^6$  m/s\cite{scalora10,ciraci12}.

Since different descriptions of nonlocality exist in the literature,
we have compared our approach to the other descriptions available
(different boundary conditions, as well as considering that the whole
response of the medium is nonlocal or not, as described in section
\ref{others} and summarize in table \ref{tab:desc}).

The hydrodynamic model is often said to exaggerate nonlocal
effects. It could be expected that taking into account the response of
the bound electrons, which can be considered as local, would lower the
impact of nonlocality on the guided mode compared to when the whole
response of the metal is considered nonlocal. Figures \ref{fig:mode}
and \ref{fig:log} show that this is paradoxically not the case when
the boundary condition that we consider as being the most physical
(${P_f}_z=0$) is not chosen. Considering a separate response of the
bound electrons actually lowers the effective plasma frequency, as
explained above, which leads to a deeper penetration of the field
corresponding to the bulk plasmons, which may in turn increase the
importance of this field (depending, of course, on the boundary
conditions).

When the condition we propose is used, the impact of nonlocality is
even lower than when considering a completely nonlocal response of the
metal and using $P_z=0$. This actually makes us think the boundary
condition we propose here, is not only the most sound physically, but
it may even yield an more accurate estimate of the nonlocal effects.

\section{Cavity resonances for metallic strips coupled to a metallic film\label{abs}}

\begin{figure}[h]
\begin{center}
\includegraphics[width=8cm]{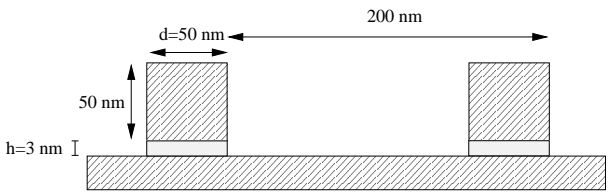}
\end{center}
\caption{Strips (rods with a 50 nm by 50 nm section) are separated from the metallic film by a 3 nm thick dielectric with an optical index of 1.58. The structure considered here is periodic, with a 200 nm period.\label{fig:schema2}}
\end{figure}

Many structures and phenomena rely on the fundamental mode of the
metallic waveguide like the enhanced transmission by subwavelength
slit arrays\cite{cao02,moreau07,collin07}, highly absorbent
gratings\cite{leperchec08} or strip
nanoantennas\cite{bozhevolnyi07,sondergaard08,jung09,yang12} to mention a
few. The latter are patches that are invariant perpendicularly to the
plane (see Fig. \ref{fig:schema2}).  The mode that is guided between a
strip and the metallic film, whose dispersion relation is given by
\eqref{eq:sym} as long as the patch is thick enough, if reflected by
the edges of the strip.  The reflection coefficient $r$ of the mode
can be computed easily\cite{moreau07} using a Fourier Modal
Method\cite{lalanne96,granet96}. When the strip is wide enough,
Fabry-Perot resonances may occur\cite{jung09}.

The local energy density (and hence the absorption) should be
proportional to the square of the field amplitude, given by a
Fabry-Perot formula\cite{moreau07}, yielding
\begin{equation}
\left|\V{H}\right|^2=\left|\frac{1}{1-r^2 \,e^{2i\,k_x\,d}}\right|^2.\label{eq:fp}
\end{equation}

\begin{figure}[h]
\begin{center}
\includegraphics[width=8cm]{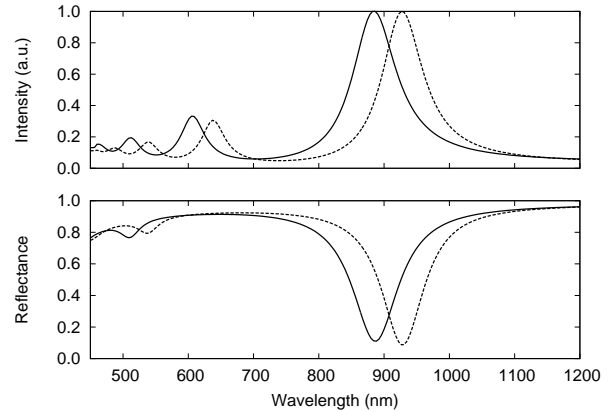}
\end{center}
\caption{Bottom : Reflection spectrum according to local (RCWA, dashed
  line) and nonlocal (COMSOL, solid line) simulations. Top : Model
  \eqref{eq:fp} for the field intensity under the strips for the local
  (dashed line) and nonlocal (solid line) theory. The agreement
  between the simulations and the model are in excellent agreement for
  the local as well as for the nonlocal theory.\label{fig:strips}}
\end{figure}

This model allows the accurate prediction of the position of the
resonance when a purely local response of the metal is assumed, as
shown figure \ref{fig:strips}: the resonance predicted using model
\eqref{eq:fp} ($k_x$ being computed using dispersion relation
Eq. \eqref{eq:sym} with $\Omega=0$) occurs exactly where there is a
dip in the reflectance of the nanorods covered surface. This confirms
the physical analysis of the structure and that a one-mode model is
sufficient to describe the resonances.

As we have shown above, when nonlocality is taken into account, the
propagation constant $k_x$ of the guided mode differs from the purely
local case. That is why the resonances of the nanorods can be expected
to be very sensitive to nonlocality when the thickness of the spacer
is typically smaller than 5 nm.

Full COMSOL simulations based on the hydrodynamical model with the boundary conditions we suggest
in this work (description 5 in table \ref{tab:desc}) show that the resonance of the structure is
largely blueshifted compared to the purely local simulations. This is completely accounted for 
by model \eqref{eq:fp} when using a propagation constant $k_x$ computed using the dispersion
relation \eqref{eq:sym} and keeping the same coefficient reflection $r$ as for the local case.
This proves that nonlocality intervenes almost only through
the change of the propagation constant of the guided mode, 
and not at all through a change of the reflection coefficient.

Such structures, or structures presenting a very similar behaviour\cite{moreau12}, 
are obviously a way to assess experimentally the effects of nonlocality on the guided
mode of the metallic waveguide with a good accuracy. 

\section{Conclusion}

We have proposed in this work an improvement of the hydrodynamic model
by clearly separating the nonlocal response of the free electrons, and
the response of the bound electrons, considered as
local\cite{liebsch1995influence}.  Such a distinction makes the
discussion about the additional boundary conditions much more clear,
leaving nothing but a single condition that seems physically sound: no
current {\em of free electrons} leaving the metal. We have shown that
this condition leads to a lower impact of the nonlocality than many
other descriptions based on the hydrodynamic model. This description
may thus answer two main concerns regarding this kind of models
compared to Feibelman's approach\cite{feibelman82,wang11}: the
uncertainty about the boundary conditions and a tendency to exagerate
the effects of nonlocality.  Furthermore, recent experimental results
have shown that the hydrodynamical model can describe nonlocal effects
very accurately\cite{ciraci12}. Given the reduced complexity of the
hydrodynamic model relative to full quantum and other microscopic
models of electron response, it is of continued interest to further
explore the accuracy of these models in the context of plasmonic
nanostructures.

Following previous work on slot waveguides that support gap
plasmons\cite{wang07}, we have shown that the slow light regime
reached when the waveguide is only a few nanometers thick is
responsible for a large enhancement of the nonlocal effects. Using
these results, we have studied the impact of nonlocality on patch
nanoantennas and shown that it should be easy to detect, paving the
way for future experiments.

Our analysis is of relevance to numerous nanophotonic devices,
including metallodielectric waveguides, nanoantennas and nanocavities,
which rely on the excitation of gap plasmons on very thin, conducting
layers for their operation\cite{miyazaki2006controlled,bozhevolnyi07,
  sondergaard08,jung09,yang12}. These resonant structures have a variety of
diverse applications, for instance, as highly efficient concentrators
and absorbers of light\cite{leperchec08,moreau12}. The description of
conductors we provide can also prove useful when testing the limits of
the classical theory for describing structures containing metals or
doped semiconductors, for which the response of the bound electrons
are strong.

As has been once more shown here, the hydrodynamic model yields
analytical results that help to understand the underlying physics of
nonlocality\cite{fuchs1981dynamical,ruppin05}. It presents the
supplementary avantage of being easy to use in simulations with
complex geometries\cite{ciraci12}. The analytical calculations we
have presented, beyond the clarification they may
bring\cite{mcmahon09,mcmahon10}, are thus a first step towards the
extension of widely used numerical
methods\cite{granet96,lalanne96,krayzel10} to account accurately for
nonlocality.

\section*{Appendix I}

Let us write equations \eqref{eq:M1} and \eqref{eq:M2} within the metal in Cartesian coordinates in the case where the fields do not depend on $y$ :

\begin{align}
-\partial_z E_y &= i\omega\,\mu_0 \,H_x\\ 
\partial_z E_x -\partial_x E_z &= i\omega\,\mu_0 \,H_y\\ 
\partial_x E_y &=i\omega\,\mu_0 \,H_z\\
-\partial_z H_y &=-i\omega\epsilon_0\epsilon \left(E_x -\alpha \partial_x^2 E_x -\alpha \partial_x\partial_z E_z \right)\\
\partial_z H_x - \partial_x H_z &=-i\omega\epsilon_0\epsilon E_y\label{Py}\\
\partial_x H_y &=-i\omega\epsilon_0\epsilon \left(E_x -\alpha \partial_z^2 E_z -\alpha \partial_x\partial_z E_x \right)
\end{align}

This system of equations can be split into two subsystems
corresponding to $s$ (electric field polarized perpendicular to the
plane of incidence) and $p$ (magnetic field polarized perpendicular to
the plane of incidence) polarizations. The $s$ subsystem is identical
to the subsystem without taking nonlocality into account, because of
the simple form of equation \eqref{Py}. Nonlocality has then no impact
on this polarization, so that we will deal in the following with $p$
polarization only.
 
The subsystem concerning the $p$ polarization can be written
\begin{align}
\partial_z E_x -\partial_x E_z &= i\omega\,\mu_0 \,H_y\label{eq:rot}\\ 
-\alpha \partial^2_x E_x + E_x-\alpha \partial_x\partial_z E_z&=\frac{1}{i\omega \epsilon_0\,\epsilon}\partial_z H_y\label{eq:a1}\\
-\alpha \partial^2_z E_z + E_z-\alpha \partial_x\partial_z E_x&=-\frac{1}{i\omega \epsilon_0\,\epsilon}\partial_x H_y\label{eq:a2}
\end{align}

By applying the operator $-\alpha \partial_x\partial_z$ to equation \eqref{eq:a2}, operator $1-\alpha \partial^2_z$ to equation \eqref{eq:a1} and subtracting one resulting equation from the other, one gets 
\begin{equation}
\left(1-\alpha \left(\partial^2_x +\partial^2_z\right)\right) E_x=\frac{\left(1-\alpha \left(\partial^2_x +\partial^2_z\right)\right)}{i\omega \epsilon_0\,\epsilon} \partial_z H_y.
\end{equation}

Repeating the same procedure, but applying $1-\alpha \partial^2_x$ to equation \eqref{eq:a2}, $-\alpha \partial_x\partial_z$ to equation \eqref{eq:a1}, and subtracting the resulting equation from the other we obtain the decoupled system of equations
\begin{align}\label{eq:vrai}
\partial_z E_x -\partial_x E_z &= i\omega\,\mu_0 \,H_y\\ 
\left(1-\alpha \left(\partial^2_x +\partial^2_z\right)\right) E_x&=\frac{\left(1-\alpha \left(\partial^2_x +\partial^2_z\right)\right)}{i\omega \epsilon_0\,\epsilon} \partial_z H_y\\
\left(1-\alpha \left(\partial^2_x +\partial^2_z\right)\right) E_z&=-\frac{\left(1-\alpha \left(\partial^2_x +\partial^2_z\right)\right)}{i\omega \epsilon_0\,\epsilon}\partial_x H_y.
\end{align}

We can apply the inverse of the differential operator $1-\alpha \left(\partial^2_x +\partial^2_z\right)$ to both sides of the equations to obtain the classical system
\begin{align}
\partial_z E_x -\partial_x E_z &= i\omega\,\mu_0 \,H_y\\ 
E_x &= \frac{1}{i\omega \epsilon_0\,\epsilon}\partial_z H_y\\
E_z &=-\frac{1}{i\omega \epsilon_0\,\epsilon}\partial_x H_y.
\end{align}
This system is identical with that corresponding to a purely local response of the metal and its solution satisfies $\Div \V{E}=0$. The wave that it describes is referred to as the {\em transverse} wave because when it is propagative, the electric field is orthogonal to the propagation vector.

But to this solution should be added any solution for which
\begin{align}
-\alpha \left(\partial^2_x +\partial^2_z\right) E_x+E_x&=0\label{long1}\\
-\alpha \left(\partial^2_x +\partial^2_z\right) E_z+E_z&=0\\
-\alpha \left(\partial^2_x +\partial^2_z\right) H_y+H_y&=0.
\end{align}
because it would also be a solution of system \ref{eq:vrai}. Using \eqref{long1} along with \eqref{eq:a1} and \eqref{eq:rot}, it is not difficult to show that $H_y=0$ so that this solution satisfies
\begin{equation}
\partial_z E_x =\partial_x E_z \label{nulcurl}.
\end{equation}

\bibliography{nonlocal}

\end{document}